\documentclass[letter,scriptaddress,twocolumn, prl,showkeys,showpacs,superscriptaddress]{revtex4-1}
\usepackage[utf8]{inputenc}
\usepackage{bm}
\usepackage{graphicx, xcolor}
\usepackage{pifont}

\begin{document}

\title{Geometric Signatures of Switching Behavior in Mechanobiology}
\author{Casey O. Barkan}
\author{Robijn F. Bruinsma}
\affiliation{Department of Physics and Astronomy, University of California, Los Angeles, Los Angeles, CA 90095}
\date{January 10, 2023}

\begin{abstract}The proteins involved in cells’ mechanobiological processes have evolved specialized and surprising responses to applied forces. Biochemical transformations that show catch-to-slip switching and force-induced pathway switching serve important functions in cell adhesion, mechano-sensing and signaling, and protein folding. We show that these switching behaviors are generated by singularities in the flow field that describes force-induced deformation of bound and transition states. These singularities allow for a complete characterization of switching mechanisms in 2-dimensional (2D) free energy landscapes, and provide a path toward elucidating novel forms of switching in higher dimensional models. Remarkably, the singularity that generates a catch-slip switch occurs in almost every 2D free energy landscape, implying that almost any bond admitting a 2D model will exhibit catch-slip behavior under appropriate force. We apply our analysis to models of P-selectin and antigen extraction to illustrate how these singularities provide an intuitive framework for explaining known behaviors and predicting new behaviors.
\end{abstract}

\maketitle

At the molecular level, mechanobiology involves a wide range of mechanical interactions between proteins that mediate cells' internal processes and their interactions with their surroundings \cite{stirnemann2022recent}. These proteins have evolved to respond to applied force in specialized and counter-intuitive ways. Non-covalent bonds that become \textit{stronger} under an applied pulling force have been found in diverse biological contexts, from cell adhesion and signaling \cite{sokurenko2008catch,mcever2015selectins, chakrabarti2017phenomenological, zhu2019dynamic}, to molecular motors \cite{guo2006mechanics, leidel2012measuring, rai2013molecular, nord2017catch}, proofreading \cite{brockman2019mechanical} and antigen discrimination \cite{zhu2019dynamic,choi2022catch, knevzevic2018active, jiang2022immune,jiang2022molecular}. Such bonds show \textit{catch} behavior (bond lifetime increases with force) over some range of forces and \textit{slip} behavior (bond lifetime decreases with force) over some other range of forces. The precise nature of the switch from catch to slip (e.g. the force at which this switch occurs) can be critical to the bond's biological function \cite{guo2006mechanics,wu2019mechano}, and one expects this switch to be tuned through evolution.

Another form of force-induced switching appears when a transformation can occur via multiple pathways: one pathway may be energetically favorable at low force, while another is favorable at higher force. This alternative form of switching also serves important functions (e.g. antigen extraction \cite{spillane2018mechanics, wang2021naturally, jiang2022immune,jiang2022molecular} and protein folding under force \cite{graham2011force, guinn2015single,wales2012evolution, pierse2017distinguishing, jagannathan2012direct}). Pathway switching and catch-slip switching often appear together in the literature \cite{makarov2016perspective, zhuravlev2016force,suzuki2011biomolecules}, in part because pathway switching can (though does not necessarily) generate a catch-slip switch \cite{bartolo2002dynamic,evans2004mechanical, pereverzev2005two}. This suggests the possibility of a unified theory for which catch-slip and pathway switching are special cases.

Early conceptual and phenomenological models of catch bonds (e.g. the two-state model \cite{barsegov2005dynamics, chakrabarti2017phenomenological}, two-pathway model \cite{evans2004mechanical, pereverzev2005two}, allosteric model \cite{sokurenko2008catch}, sliding-rebinding model \cite{lou2007structure}) have had success explaining many experimental observations. More recent theoretical considerations have revealed that multi-dimensionality of the bond's free energy landscape is necessary for catch-slip behavior \cite{zhuravlev2016force,suzuki2010single,makarov2016perspective}. In particular, the deformation of bound state and transition state (i.e. the movement of minimum and saddle points through configuration space) under applied force can generate a variety of catch-slip behaviors \cite{suzuki2011biomolecules,konda2013molecular,quapp2016reaction,chakrabarti2014plasticity,adhikari2018unraveling} and force-history dependence \cite{kong2013cyclic,chen2015probing,li2016model} in simple 2-dimensional course-grained free energy landscapes. With many 2-dimensional models of catch-slip behavior now known, we are led to ask: how generic is this switching behavior? Are there features of free energy landscapes that indicate a switch?

We discover geometric signatures of catch-slip and pathway switching behavior in the form of singularities in the flow field that describes force-induced deformation of minimum and saddle points. These singularities, which we call ``switch points", can be viewed as the basic building blocks of force-induced switching behavior in two dimensional systems. Using this framework, we show that virtually every 2-dimensional bond will exhibit catch-slip behavior under an appropriate force and/or stress. In higher dimensions, the generalization of switch points indicate other, more exotic, responses to force.

Switch points provide a unified view of known switching mechanisms, and we show how established catch bond models can be understood using switch points. Furthermore, switch points serve as a guide to elucidate new possibilities and make new predictions. We illustrate their utility with a model of the P-Selectin catch bond and a generalized model of the `tug-of-war' process in which B-cells probe antigen specificity via a pathway switch. 



\textit{Bond rupture under applied force} --- Consider a system described by a two-dimensional vector $\bm x$ governed by a free energy landscape (or potential of mean force) $V(\bm x)$. The external force couples linearly to $\bm x$ along a direction $\bm{\hat \ell}$ and with magnitude $f$. We suppose that the coupling direction $\bm{\hat \ell}$ is fixed while the magnitude $f$ is varied, so that the total potential is
\begin{equation}
    V_f(\bm x)=V(\bm x)-f\bm{\hat \ell}\cdot \bm x
\end{equation}
A potential $V_f$ describing a meta-stable bond must have a local minimum corresponding to the bonded state, and one or more saddle points which correspond to pathways along which bond rupture can occur. For a single pathway, the bond lifetime $\tau$ can be approximated using Langer's formula $\tau=\nu\exp{(E_b/k_BT)}$ where the energy barrier $E_b$ is the difference in $V_f$ between the minimum and saddle and the prefactor $\nu$ captures entropic effects of the minimum and saddle \cite{langer1969statistical}. When there are multiple pathways, the lifetime is approximately $\tau=(\sum_i\tau_i^{-1})^{-1}$ where $\tau_i$, the mean first passage time over pathway $i$, is given by Langer's formula.

As the force magnitude $f$ is varied, the minimum and saddle points of $V_f$ move through the configuration space. In other words, the bound and transition state(s) are deformed by the force,  causing a force-dependence of $E_b(f)$ and $\nu(f)$ in Langer's formula. The force-dependence of $\tau(f)$ is typically dominated by the force-dependence of $E_b(f)$ \cite{suzuki2010single, avdoshenko2016reaction}, so $E_b(f)$ will be our primary focus. For one dimensional systems, a pulling force always causes the minimum and saddle to move toward each other, decreasing $E_b$ (slip bond behavior) \cite{zhuravlev2016force}. However, in two or more dimensions, the minimum and saddle(s) may take complicated paths through configuration space as $f$ is increased, leading to catch-slip and pathway switches \cite{suzuki2011biomolecules,konda2013molecular,quapp2016reaction,quapp2018toward}. Fig.~1A and B show examples of movements of a minimum (\ding{108}) and one or two saddles (\ding{54}) under increasing $f$ generating, respectively, catch-slip via a single pathway and switching between pathways. Figure ~1C and D show the energy barrier vs. $f$ corresponding to Fig.~1A and B, respectively. Vertical bars indicate where each switch occurs. The movement of any critical point $\bm x_c$ (minimum, saddle, or maximum) obeys \cite{konda2014exploring,quapp2016reaction}
\begin{equation}\label{dxdf}
    \frac{d}{df} \bm x_c = H^{-1}(\bm x_c)\bm{\hat \ell}
\end{equation}
where $H(\bm x_c)$ is the Hessian matrix of $V(\bm x)$ at $\bm x_c$. Importantly, Eq.~\ref{dxdf} has no explicit $f$ dependence, so it defines an autonomous dynamical system in the parameter $f$. The initial condition of this dynamical system can be adjusted by applying a constant force to the bond (we expand on this point below). Solutions of Eq.~\ref{dxdf} can be used in Langer's formula to find the force-dependent bond lifetime $\tau(f)$, which, given a time-dependent force protocol $f(t)$, provides the experimentally-measurable survival probability $p(t)=\exp[-\int_0^tdt'\;\tau(f(t'))^{-1}]$ \cite{evans1997dynamic,dudko2006intrinsic}.

\begin{figure}
\includegraphics[width=8.4cm]{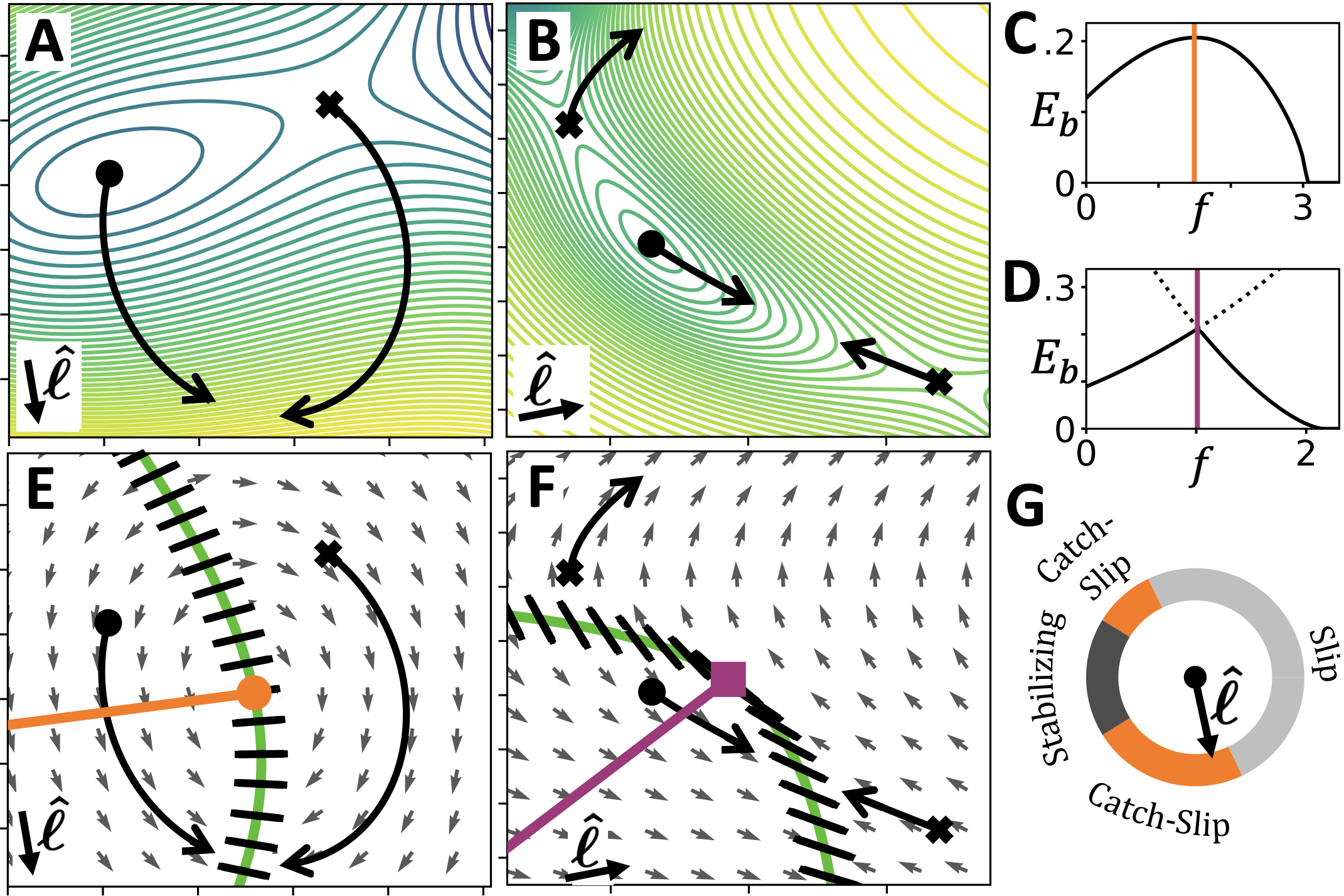}
\caption{\label{fig:1} 
\textbf{Switch points are signatures of switching behavior}. \textbf{A} and \textbf{B}: Two examples showing paths of minimum (\ding{108}) and saddle (\ding{54}) under increasing force (i.e. solutions of Eq.~\ref{dxdf}). Contour lines show $V_f(\bm x)$ at $f=0$. Color scheme: blue-to-yellow indicates low-to-high $V(\bm x)$. Note that panel B is the classic `two-pathway' model \cite{evans2004mechanical,sokurenko2008catch}. \textbf{C} and \textbf{D}: Energy barrier $E_b(f)$ corresponding to panels A and B, respectively. In panel D, solid line shows $E_b$ over the lower-energy saddle and dashed line shows $E_b$ over the higher-energy saddle. Vertical bars indicate the occurrence of a catch-slip switch via a single pathway (panel C) and a switch in preferred pathway (panel D). \textbf{E} and \textbf{F}: Flow vector field, $\det(H){=}0$ curve (green) with attached $\bm{\hat v}_0$ vectors (black hash marks) corresponding to panels A and B, respectively. Orange dot and purple square indicate $\bm{\hat \ell}$- and $\bm{\hat n}$-switch points. Orange and purple lines indicate switch lines. \textbf{G:} Dial indicating behavior of the bond in panel A for all possible force directions $\hat{\bm\ell}$. The direction of $\hat{\bm\ell}$ from panel A is indicated.}
\end{figure}

\textit{Switch points} --- The flow described by Eq.~\ref{dxdf} can have singularities that generate a force-induced switch. To see this, first note that Eq.~\ref{dxdf} is undefined on the curve where $\det(H){=}0$, where $H$ is the $\bm x$-dependent Hessian. This $\det(H){=}0$ curve separates regions of configuration space with minimum-like curvature from regions with saddle-like curvature, from regions with maximum-like curvature. The $\det(H){=}0$ curve behaves as either a source or a sink under the flow. Fig.~1E and F show the flow vector field and $\det(H){=}0$ curve (green) corresponding to Fig.~1A and B, respectively. As shown, the flow points either outward away from the $\det(H){=}0$ curve (source) or inward towards it (sink). Hence, solutions of Eq.~\ref{dxdf} (i.e. integral curves of the flow) emerge from and/or coalesce with the $\det(H){=}0$ curve. As a consequence, \textit{global} behavior of Eq.~\ref{dxdf} is revealed by examining \textit{local} behavior near the $\det(H){=}0$ curve. This local behavior is described by the equation (see SM)
\begin{equation}\label{dxdf_approx}
    \frac{d}{df} \bm x_c \propto (\bm{\hat{\ell}}\cdot \bm{\hat{v}_0}) \bm{\hat{v}_0} + \mathcal{O}(\lambda_0)
\end{equation}
Here, $\lambda_0$ is the eigenvalue of $H$ that passes through zero along the $\det(H){=}0$ curve, and $\bm{\hat{v}_0}$ is the corresponding eigenvector. Importantly, $\bm{\hat{v}_0}$ and $\lambda_0$ (which are functions of position along the $\det(H){=}0$ curve) are independent of applied force, as is the $\det(H){=}0$ curve itself. Eq.~\ref{dxdf_approx} implies that flow vectors along the $\det(H){=}0$ curve point outward or inward parallel to $\bm{\hat{v}_0}$. Letting $\bm{\hat n}$ denote an outward-pointing unit vector normal to the $\det(H){=}0$ curve, the flow vectors point outward if $\frac{d}{df}\bm x_c\cdot\bm{\hat n}>0$, or equivalently, $(\bm{\hat \ell}\cdot \bm{\hat v_0})(\bm{\hat v_0} \cdot \bm{\hat n})>0$. They point inward if $(\bm{\hat \ell}\cdot \bm{\hat v_0})(\bm{\hat v_0} \cdot \bm{\hat n})<0$. We define a \textit{switch point} as a singular point at which the quantity $(\bm{\hat \ell}\cdot \bm{\hat v_0})(\bm{\hat v_0} \cdot \bm{\hat n})$ passes through zero. In other words, switch points are where the flow switches from outward (source) to inward (sink). Switch points come in two varieties: 
$\bm{\hat\ell}$-switch points, where $\bm{\hat \ell}\cdot \bm{\hat v_0}=0$, and $\bm{\hat n}$-switch points, where $\bm{\hat n}\cdot \bm{\hat v_0}=0$. As suggested from the examples in Fig.~1, $\bm{\hat\ell}$-switch points signify a catch-slip (or slip-catch) switch via a single pathway, and $\bm{\hat n}$-switch points signify a switch between pathways. In the SM we discuss the geometric basis for these behaviors of $\bm{\hat\ell}$- and $\bm{\hat n}$- switch points. To identify the location of switch points, one can simply plot the $\bm{\hat v_0}$ vectors along the $\det(H){=}0$ curve. These are shown as black hash marks in Fig.~1D and F. Note that the sign of $\bm{\hat v_0}$ is immaterial, so the hash marks have no arrow to indicate a sign. $\bm{\hat\ell}$-switch points are located where a hash mark is perpendicular to $\bm{\hat\ell}$ (i.e. $\bm{\hat \ell}\cdot \bm{\hat v_0}=0$), and $\bm{\hat n}$-switch points are located where a hash mark is tangent to the $\det(H){=}0$ curve (i.e. $\bm{\hat n}\cdot \bm{\hat v_0}=0$). The switch points in Fig.~1E and F are shown as an orange dot and purple square, respectively.

Emanating from a switch point is a \textit{switch line} (orange and purple lines in Fig.~1E and F, respectively) which marks where the switch occurs. Specifically, the switch in behavior occurs when the minimum crosses the switch line under increasing $f$. A switch line emanating from an $\bm{\hat\ell}$-switch point indicates catch-slip or slip-catch via a single pathway:  $\frac{d}{df}E_b=0$ when the minimum is on the switch line. A switch line emanating from an $\bm{\hat n}$-switch point indicates a switch in pathway, i.e. the lowest-energy saddle switches from one saddle to another as the minimum crosses the switch line. For a minimum on this type of switch line, the energy barriers of two saddles are equal. Switch lines partition configuration space into regions of a given behavior. For instance, in Fig.~1E, the minimum begins in a `catch region' and is pulled across the switch line into a `slip region' as $f$ increases. To third order in $V(\bm x)$, switch lines are straight lines, and equations for their direction are given in the SM.

The framework of switch points and switch lines reveals behaviors that would otherwise be `invisible' from viewing the free energy landscape at any particular force. Three examples are:

\textbf{\textit{(i)}} Consider modifying the force direction $\bm{\hat \ell}$ in Fig.~1A and E. Varying $\bm{\hat \ell}$ moves the $\bm{\hat\ell}$-switch point along the $\det(H){=}0$ curve, modifying the force at which the switch from catch to slip occurs. Fig.~1G shows the behavior for any given direction $\bm{\hat \ell}$. Slip behavior occurs for a wide range of directions that are roughly aligned with the reaction pathway of bond rupture (i.e the direct path from minimum to saddle). For force directions opposite to the reaction pathway, the force stabilizes the bond, acting as a ``pushing" force rather than a ``pulling" force. Between the slip regime and stabilizing regime are catch-slip regimes. As indicated, the $\bm{\hat\ell}$ in Fig.~1A and E falls into the catch-slip regime. 

\textbf{\textit{(ii)}} Consider that the bond may be under stress due to its local environment, resulting in an additional force $\bm h$ on the bond which moves the initial condition of Eq.~\ref{dxdf} to points satisfying $\nabla V=\bm h$. The force $\bm h$ can move the initial position of the minimum across a switch line, from a catch region to a slip region, or vice versa. This phenomenon may be at play in two-site `dynamic catch' bonds \cite{fiore2014dynamic,zhu2019dynamic}, where a ligand with two binding sites exhibits slip behavior when bound to either site individually, but catch-slip behavior when both sites are bound. We analyze a toy model of such a scenario in the SM. Briefly summarized, if the ligand is bound at just one site, the minimum is in a slip region. When the ligand binds at the second site, the minimum is pulled into the catch region, so that the system exhibits catch-slip behavior only when both sites are bound. 

\textbf{\textit{(iii)}} Whether or not a pathway switch may occur cannot be inferred simply from the existence or absence of multiple saddle points because pathways can be created or destroyed as $f$ varies \cite{konda2014exploring}. However, the existence (or absence) of an $\bm{\hat n}$-switch point is a definitive indication of the possibility (or impossibility) of a pathway switch.

The framework of switch points also lets us address the question: how prevalent are free energy landscapes that exhibit catch-slip behavior? Remarkably, the answer is that \textit{almost every 2-dimensional $V(\bm x)$ will exhibit catch-slip behavior for an appropriate $\bm{\hat\ell}$ and $\bm h$}. Indeed, there exists an $\bm{\hat\ell}$ that generates an $\bm{\hat\ell}$-switch point for all but a measure-zero subset of smooth functions $V(\bm x)$ (the only requirement for the existence of such an $\bm{\hat\ell}$ is that not all $\bm{\hat v_0}$ be parallel along the $\det(H){=}0$ curve separating minimum-like and saddle-like regions). Given these observations, we suggest that investigating pulling directions and internal stresses, rather than investigating complex bonding mechanisms, may be the key to understanding much of the catch-slip behavior observed in biological systems.

\textit{Application to P-Selectin} --- To illustrate the utility of the geometric framework of switch points and switch lines, we develop a model of the catch-slip bond between P-Selectin and its glycoprotein ligand-1 (PSGL-1). This catch-slip behavior is believed to mediate leukocyte `rolling' \cite{mcever2010rolling}. We propose a 2-dimensional free energy landscape for this bond based on structural considerations and we find close agreement with experimental data \cite{marshall2003direct} with just with 3 fit parameters (see SM). Fig.~2A shows the structure of P-Selectin (teal) and PSGL-1 (magenta) and the two degrees of freedom: angle $\theta$ between two domains and distance $d$ between protein and ligand.

\begin{figure}
\includegraphics[width=8.5cm]{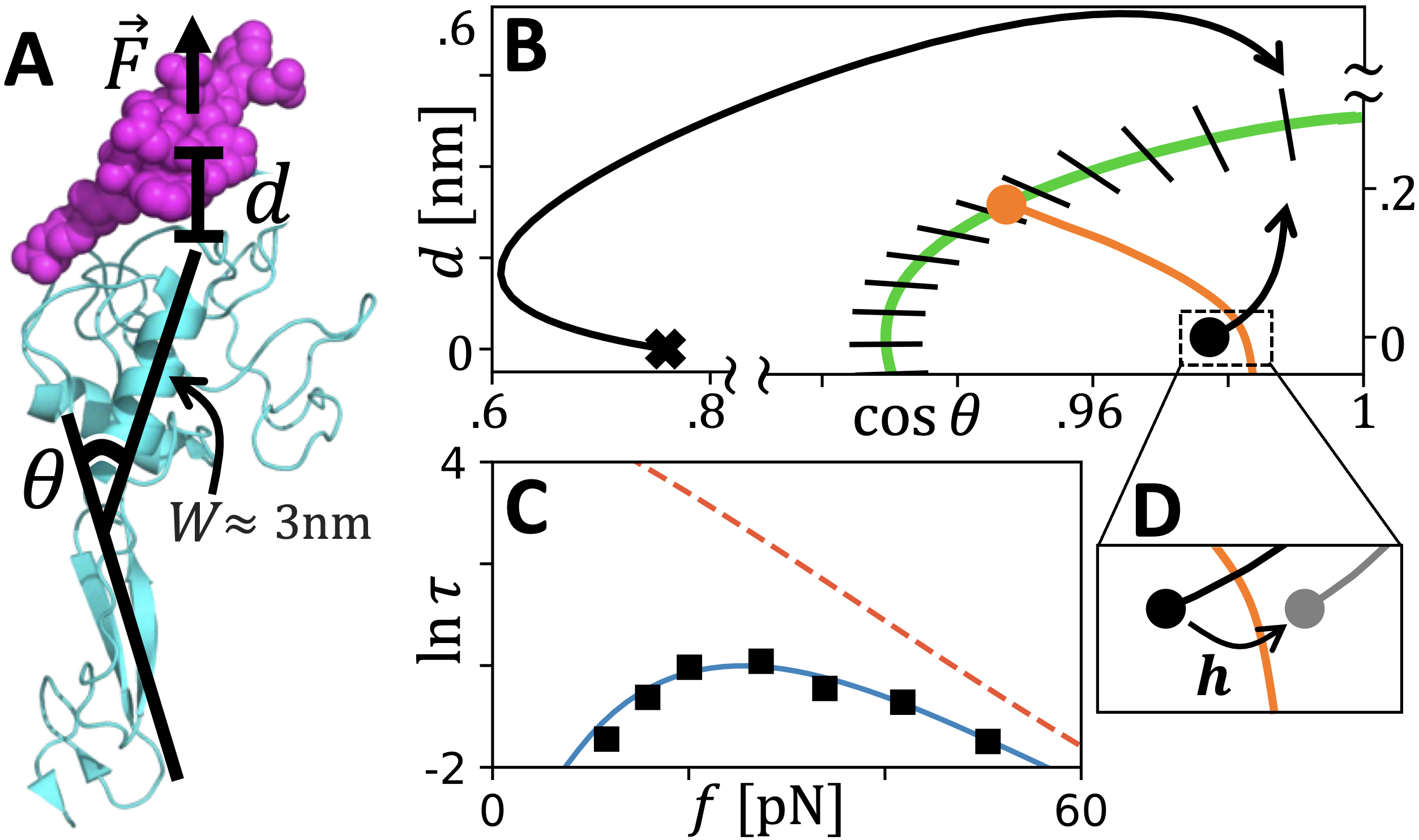}\caption{\textbf{Model of P-Selectin contains $\bm{\hat \ell}$-switch point.} \textbf{A}: Structure of P-Selectin--PSGL-1 (PDB ID code: 1G1S \cite{somers2000insights}, rendered in PyMOL) with two collective degrees of freedom $\theta$ and $d$. \textbf{B}: Geometric framework reveals $\bm{\hat \ell}$-switch point (orange dot) resulting from force direction $\bm{\vec\ell}=(W,1)$, $W{\approx}3\textrm{nm}$ as indicated. $\bm{\vec\ell}$ is set by the structure of the model (see SM). \textbf{C}: log of $\tau$ (seconds) vs. force (pN) from experiment (black squares \cite{marshall2003direct}), from our fitted model (solid blue), and from our prediction of bond behavior under the additional (generalized) force $\bm h=70\textrm{pNnm} \bm{\hat x_1}$ where $\bm{\hat x_1}$ is the direction of increasing $\cos\theta$ (dashed orange). \textbf{D}: Displacement of minimum due to $\bm h$.}
\end{figure}

The geometric framework (Fig.~2B) reveals one $\bm{\hat \ell}$-switch point (orange dot) and no $\bm{\hat n}$-switch points---this is topologically equivalent to the example in Fig.~1E. The absence of an $\bm{\hat n}$-switch point indicates that a pathway switch cannot occur in this model at any force, regardless of mangitude or direction (recall that the absence of two saddles at any \textit{particular} force is insufficient to make this strong claim). Under increasing force, the minimum is pulled across the switch line (Fig.~2B orange curve), generating catch-slip behavior with force-dependent bond lifetime $\tau(f)$ in close agreement with experimental data \cite{marshall2003direct} (Fig.~2C; experiment: black squares; model: solid blue curve). Our model predicts that an additional force $\bm h$ pushing upward on the lectin domain will move the minimum across the switch line (Fig.~2D), erasing the catch behavior as well as strengthening the bond at zero force (Fig.~2C dashed orange curve shows the predicted $\tau(f)$). In the SM we discuss this prediction in more detail and speculate that this response may put an evolutionary constraint on the length of the consensus repeats \cite{mcever2010rolling} that attach the lectin domain to the membrane.

\textit{Interplay of $\bm{\hat \ell}$- and $\bm{\hat n}$-switch points} --- $\bm{\hat \ell}$- and $\bm{\hat n}$-switch points can appear in the same bond, generating catch-slip and pathway switching at different forces. As an example, we generalize a recently-proposed model for the ``tug-of-war" process of antigen (Ag) extraction by B-cell receptors (BCR) \cite{jiang2022immune, jiang2022molecular}. Fig.~3A shows a schematic of this system: a BCR (blue) and antigen-presenting cell receptor (APC, green) pull on either end of an Ag fragment (orange) until rupture occurs either by pathway 1 or 2 (as labelled). A force-induced switch in this rupture pathway is believed to signal antigen affinity. While the original model treats the BCR-Ag and APC-Ag bonds as independent \cite{jiang2022immune}, coupling between these bonds (which could arise, for example, from allistory of the antigen) generates a richer geometric picture (Fig.~3B). An $\bm{\hat n}$-switch point (purple square) and associated switch line indicates where the pathway switch occurs. Additionally, an $\bm{\hat \ell}$-switch point (orange dot) exists due to the force direction $\bm{\hat \ell}$ imposed by the geometry (see SM). This results in slip-catch behavior via pathway 1 at low forces, so that the BCR-Ag bond is strong at zero force, easily dissociates at intermediate force, then strengthens at higher force so that rupture via the APC-Ag bond becomes favorable. The energy barrier vs. $f$ is shown in Fig.~3C. The slip-catch switch at low force and pathway switch at high force are marked by orange and purple bars, respectively. Such slip-catch-slip, or \textit{triphasic} behavior has been observed in selectins \cite{chakrabarti2017phenomenological}, and in the SM we further discuss the plausibility of this behavior in antigen extraction.

\begin{figure}[b]
\includegraphics[width=8.5cm]{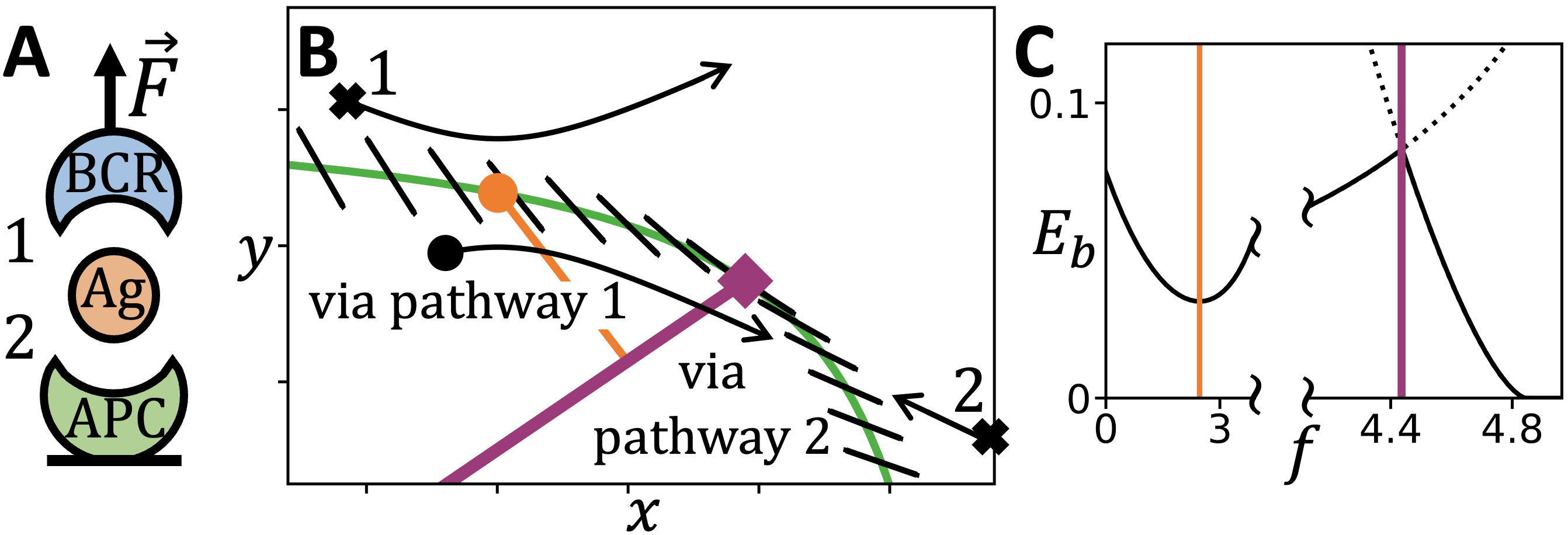}\caption{\textbf{A generalized `tug-of-war' model.} \textbf{A:} Schematic of the `tug-of-war' system \cite{jiang2022molecular}. \textbf{B:} Geometric picture reveals an $\bm{\hat\ell}$-switch point (orange dot) and $\bm{\hat n}$-switch point (purple square), generating a clip-catch switch and pathway switch, respectively.  Switch lines extending from each switch point indicate where the switch occurs. Coordinates $x$ and $y$ are the distance between APC and Ag, and between Ag and BCR, respectively. \textbf{C:} $E_b$ vs. $f$. Orange and purple vertical bars indicate crossing of slip-catch and pathway switch lines, respectively.}
\end{figure}

In more complex bonds, many $\bm{\hat \ell}$- and $\bm{\hat n}$-switch points can appear together. In the SM we analyze a model of the sliding-rebinding catch-bond mechanism \cite{lou2007structure}. This model contains four switch points in total, and our geometric framework organizes this complex scenario into an intuitive picture. Note that in 2 dimensions, $\bm{\hat \ell}$- and $\bm{\hat n}$-switch point are the only singularities that can occur in the flow along the $\det(H){=}0$ curve, so combinations of 
$\bm{\hat \ell}$- and $\bm{\hat n}$-switch points cover all possible switching behaviors in 2 dimensional models. In higher dimensions, switch points can be generalized, and we show in the SM that this generalization reveals more exotic responses to applied force, in addition to catch-slip and pathway switching.

\textit{Discussion} --- We present a geometric framework for characterizing force-induced switching behavior in 2-dimensional free energy landscapes. We find geometric signatures---switch points and switch lines---that identify both catch-slip switches via a single pathway and switching between pathways. Using this framework, we show that almost every 2-dimensional free energy landscape will exhibit catch-slip behaviour under an appropriate force. Indeed, very simple bonds show catch behavior when pulled in the right direction and/or put under certain stresses. This motivates experiments that probe multiple pulling directions (as in \cite{gittes1996directional,nicholas2015cytoplasmic,jagannathan2012direct,huang2017vinculin}) and investigations into the orientation of bonds in their native context. Additionally, the ubiquity of catch-slip behavior suggests simple bonds may serve as evolutionary stepping-stones in the evolution of specialized catch bond mechanisms. In the SM, we analyze established catch-bond models and identify the signatures of their switching behavior. 

Higher-dimensional generalizations of switch points will enlarge the applicability of this framework. In the SM, we show that such generalized switch points signify more exotic responses to force, in addition to catch-slip and pathway switches. Future work is needed to explore the range of possible behaviors. An assumption of our approach is that the force dependence of bond lifetime $\tau(f)$ is determined predominantly by the force-dependence of the energy barrier $E_b(f)$. While this is typically well justified \cite{avdoshenko2016reaction}, it was recently suggested experimentally that force-dependent entropic effects (captured by Langer's prefactor $\nu$) can be important \cite{farago2021activated}, motivating theoretical investigation into this possibility.

\begin{acknowledgments}
We thank our late friend and colleague Alex Levine for suggesting this project and Shenshen Wang for introducing us to the importance of catch bonding and pathway switching in the context of receptor-antigen interactions and also for providing us with important references. C.B. also thanks Jonathon Howard for stimulating discussions and references. C.B. is grateful for support from the NSF Graduate Research Fellowship Program (NSF Grant No. DGE-2034835) and R.B. would like to thank the NSF-DMR for continued support under CMMT Grant No.1836404.
\end{acknowledgments}

\bibliography{citations}

\end{document}